\documentclass[aps,prl,reprint,superscriptaddress,10pt]{revtex4-1}
\usepackage{amsmath,amssymb,amsfonts,graphicx,bm,color}
\usepackage{mathrsfs}
\usepackage[utf8]{inputenc}
\usepackage{textcomp}
\usepackage[T1]{fontenc}
\usepackage{lmodern}

\usepackage{ulem}

\begin{document}

\title{Monte Carlo simulations in the unconstrained ensemble}

\author{Ivan Latella}
\email{ilatella@ub.edu}
\affiliation{Departament de F\'isica de la Mat\`eria Condensada, Universitat de Barcelona, Mart\'i i Franqu\`es 1, 08028 Barcelona, Spain}

\author{Alessandro Campa} 
\affiliation{National Center for Radiation Protection and Computational Physics, Istituto Superiore di Sanit\`{a}, Viale Regina
Elena 299, 00161 Roma, Italy}

\author{Lapo Casetti}
\affiliation{Dipartimento di Fisica e Astronomia, Universit\`a di Firenze, and INFN, Sezione di Firenze,
via G. Sansone 1, 50019 Sesto Fiorentino, Italy}
\affiliation{INAF-Osservatorio Astrofisico di Arcetri, Largo E. Fermi 5, 50125 Firenze, Italy}

\author{Pierfrancesco Di Cintio}
\affiliation{Dipartimento di Fisica e Astronomia, Universit\`a di Firenze, and INFN, Sezione di Firenze,
via G. Sansone 1, 50019 Sesto Fiorentino, Italy}

\author{J. Miguel Rubi}
\affiliation{Departament de F\'isica de la Mat\`eria Condensada, Universitat de Barcelona, Mart\'i i Franqu\`es 1, 08028 Barcelona, Spain}

\author{Stefano Ruffo}
\affiliation{SISSA, via Bonomea 265 and INFN, Sezione di Trieste, 34136 Trieste, Italy}
\affiliation{Istituto dei Sistemi Complessi, Consiglio Nazionale delle Ricerche, via Madonna del Piano 10, 50019 Sesto Fiorentino, Italy}

\begin{abstract}
The unconstrained ensemble describes completely open systems whose control parameters are chemical potential, pressure, and temperature. 
For macroscopic systems with short-range interactions, thermodynamics prevents the simultaneous use of these intensive variables as control parameters, because they are not independent and cannot account for the system size. When the range of the interactions is comparable with the size of the system, however, these variables are not truly intensive and may become independent, so equilibrium states defined by the values of these parameters may exist.
Here, we derive a Monte Carlo algorithm for the unconstrained ensemble and show that simulations can be performed using chemical potential, pressure, and temperature as control parameters. We illustrate the algorithm by applying it to physical systems where either the system has long-range interactions or is confined by external conditions.
The method opens up a new avenue for the simulation of completely open systems exchanging heat, work, and matter with the environment.
\end{abstract}

\maketitle

The Metropolis Monte Carlo (MC) method~\cite{Metropolis_1953} vastly contributed to the understanding of many physical phenomena~\cite{Frenkel,Landau}.
Different versions of the method have been devised, applied to various statistical ensembles as, e.g., microcanonical~\cite{Creutz_1983}, canonical~\cite{Metropolis_1953}, grand canonical~\cite{Norman_1969}, semi-grand canonical~\cite{Kofke_1988}, isothermal-isobaric~\cite{Wood_1968,McDonald_1972}, isostress-isostrain~\cite{Schoen_1993}, and distinct variants of the Gibbs ensemble~\cite{Panagiotopoulos_1987,Panagiotopoulos_1988}.
All these ensembles include at least one extensive variable as a control parameter, such as energy, volume or number of particles: little attention has been paid to MC methods in which the control parameters are the chemical potential $\mu$, pressure $P$, and temperature $T$.
Indeed, thermodynamics tells us that the latter intensive quantities are not independent and cannot account for the size of a macroscopic system with short-range interactions~\cite{Frenkel,Callen}.
When applied to these systems, the $\mu PT$ ensemble requires the addition of the equation of state linking $\mu$, $P$, and $T$ or, conversely, it can be used to infer such a link~\cite{Hill_SM}.
Reported MC methods~\cite{Orkoulas_2009,Wilding_2013} have taken advantage of this fact by considering a constrained $\mu PT$ ensemble in which the examined systems are partially closed in either volume or number of particles, thus finding the underlying relation $\mu(P,T)$ of the system under scrutiny.

In contrast, an unconstrained ensemble with $\mu$, $P$, and $T$ as independent control parameters can be properly defined for small systems~\cite{Hill}, confined systems~\cite{Schoen_1994,footnote}, and for long-range interacting systems~\cite{Latella_2017,Campa_2020}, which are intrinsically nonadditive~\cite{Campa,Latella_2015} and have an additional degree of freedom that may render $\mu$, $P$, and $T$ independent.
This makes it possible to study completely open systems exchanging heat, work, and matter with the environment.
In this Letter we derive an elementary MC scheme for simulations in the unconstrained ensemble and show that it consistently combines the MC algorithms of the grand canonical and isothermal-isobaric ensembles.
By testing the method for simple physical systems that can be analytically evaluated, we identify the role of interactions in equilibrium states of completely open systems.

Since dealing with completely open systems is rather unconventional, we first recall some concepts about their thermostatistics~\cite{Hill,Latella_2017}.
Considering for simplicity a one-component system where the combination of the first and second laws of thermodynamics is expressed by $dE=TdS-PdV+\mu dN$, where $E$, $S$, $V$, and $N$ are the energy, entropy, volume, and number of particles, respectively, one defines a statistical ensemble by taking a collection of $\mathscr{N}$ independent replicas of the system, with total energy, entropy, volume, and number of particles $E_t=\mathscr{N} E$, $S_t=\mathscr{N} S$, $V_t=\mathscr{N} V$, and $N_t=\mathscr{N} N$, respectively.
The energy balance for the ensemble becomes~\cite{Hill}
\begin{equation}
dE_t=TdS_t-PdV_t+\mu dN_t+\mathscr{E}d\mathscr{N}, 
\label{first-law_ensemble}
\end{equation}
where the last term accounts for the variation of $E_t$ when $\mathscr{N}$ varies at constant $S_t$, $V_t$, and $N_t$; the quantity $\mathscr{E}$ is called subdivision potential~\cite{Hill} or replica energy~\cite{Latella_2015,Bedeaux}.
No assumption has been made on the nature of the system, so Eq.~(\ref{first-law_ensemble}) is general.
It can be integrated holding $E$, $S$, $V$, and $N$ constant, arriving at $\mathscr{E}= E-TS+PV-\mu N$ (see the Supplemental Material~\cite{Supplemental} for details).
The variation of $\mathscr{E}$, making use of $dE=TdS-PdV+\mu dN$, yields $d\mathscr{E}= - N d\mu + V dP -S dT$, showing that the natural variables of the replica energy are $\mu$, $P$, and $T$. 
One realizes that conventional thermodynamics~\cite{Callen} focuses on systems in which $\mathscr{E}=0$ by imposition, but this is not the general situation. 
Systems with $\mathscr{E}\neq0$ are nonadditive~\cite{Latella_2015}, because their entropy is not a linear homogeneous function of $E$, $V$, and $N$.
As discussed below, $\mathscr{E}$ can be derived~\cite{Hill} from a partition function $\Upsilon(\mu,P,T)$ such that
$\mathscr{E}(\mu,P,T)=-k_BT\ln\Upsilon(\mu,P,T)$, where $k_B$ is the Boltzmann constant.

\begin{figure}
\includegraphics[width=\columnwidth]{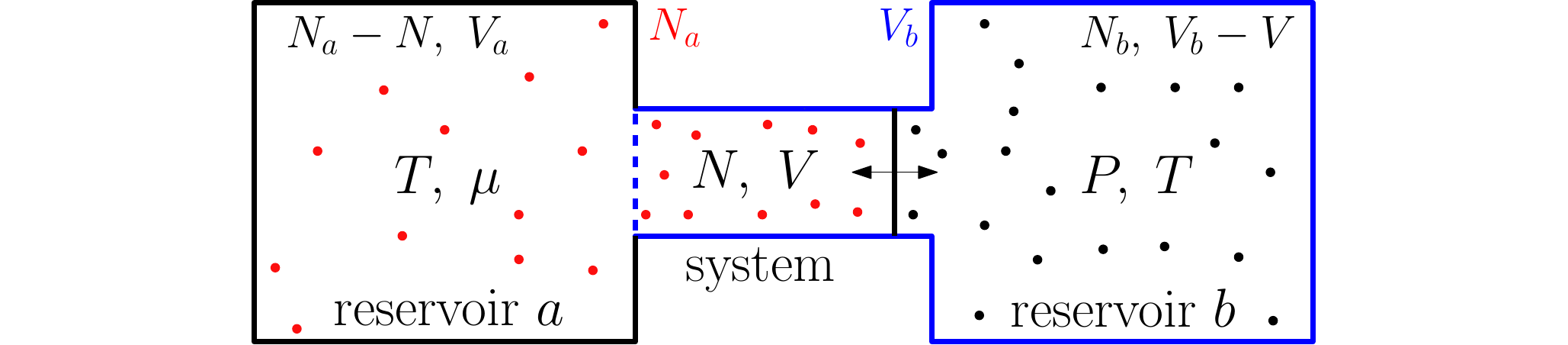}
\caption{System coupled to two reservoirs under completely open conditions.}
\label{sketch}
\end{figure}

To establish the basis for the MC algorithm in the unconstrained ensemble, we closely follow Ref.~\cite{Frenkel}, extending the standard arguments used for other ensembles.
For simplicity we consider a system in a cubic box of side $L=V^{1/3}$ (the extension to a rectangular box is immediate) in which particles have coordinates $\mathbf{r}_i$, $i=1,\dots,N$, and define scaled coordinates by
$\mathbf{r}_i=V^{1/3}\mathbf{s}_i$.
The canonical partition function of the system takes the form
\begin{equation}
Q(N,V,T)=\frac{V^N}{\Lambda^{3N}N!}\int_0^1 \cdots\int_0^1 d\mathbf{s}^N e^{-\beta\mathcal{U}(\mathbf{s}^N;V)}, 
\label{Q_canonical}
\end{equation}
where $\mathbf{s}^N \equiv (\mathbf{s}_1,\dots,\mathbf{s}_N)$, $\beta=1/k_BT$, $\Lambda$ is the de Broglie thermal wavelength, and $\mathcal{U}(\mathbf{s}^N;V)$ is the potential energy. 
We assume that the system is coupled to two independent reservoirs of ideal gas particles at temperature $T$: reservoir $a$ that exchanges particles with the system and reservoir $b$ that exchanges volume (see Fig.~\ref{sketch}). 
Reservoir $a$ has $N_a-N$ particles and volume $V_a$, reservoir $b$ has $N_b$ particles and volume $V_b-V$, and their canonical partition functions $Q_a(N_a-N,V_a,T)$ and   $Q_b(N_b,V_b-V,T)$ are easily obtained from Eq.~(\ref{Q_canonical}) by setting $\mathcal{U}=0$. 
The partition function of the total system including the reservoirs is $Q^\mathrm{tot}=Q_aQQ_b$. 
Thus, the probability density of observing the system with $N$ particles and volume $V$ is
$\mathcal{P}(N,V)=Q^\mathrm{tot}/\sum_{N=0}^{N_a}\int_0^{V_b}dVQ^\mathrm{tot}$, so the most probable values of $N$ and $V$ are those that minimize the total free energy $F^\mathrm{tot}=-k_BT\ln Q^\mathrm{tot}$.
Taking the limit of infinite reservoirs, the quantities that survive, besides $T$, are the chemical potential $\mu$ of reservoir $a$ and the pressure $P$ of reservoir $b$ (see the Supplemental Material~\cite{Supplemental} for details).
Hence, the probability density of finding the system in volume $V$ in a particular $N$-particle configuration takes the form
\begin{equation}
\mathcal{P}(N,V;\mathbf{s}^N) =\frac{\beta P e^{\beta\mu N-\beta PV+N\ln(V/\Lambda^3)-\ln N!-\beta\mathcal{U}(\mathbf{s}^N;V)} }{\Upsilon(\mu,P,T)} ,
\label{prob_unconstrained}
\end{equation}
where the unconstrained partition function is given by
\begin{equation}
\Upsilon(\mu,P,T)=\beta P\sum_{N=0}^\infty e^{\beta\mu N}\int_0^\infty dV e^{-\beta PV}Q(N,V,T).
\label{unconstrained_partition_fucntion}
\end{equation}
Notice that it would not be possible to implement completely open conditions with a single reservoir of ideal gas particles, since $\mu$, $P$, and $T$ are not independent in this case.

Given a system configuration $\mathcal{C}$ from which a new configuration $\mathcal{C}'$ is generated in the simulation, we follow the Metropolis algorithm~\cite{Frenkel} using Eq.~(\ref{prob_unconstrained}) and compute the acceptance probability of the new configuration as $P_\mathrm{acc}(\mathcal{C}\to \mathcal{C}')=\min\left[1, \mathcal{P}(\mathcal{C}')/\mathcal{P}(\mathcal{C})\right]$.
MC moves in this case consist of displacements of particles, insertion and removal of particles, and changes of volume, yielding a potential energy variation $\Delta\mathcal{U}=\mathcal{U}(\mathcal{C}')-\mathcal{U}(\mathcal{C})$.
A particle displacement is attempted by selecting a single particle at random with coordinates $\mathbf{s}$ and performing a random displacement from $\mathbf{s}$ to $\mathbf{s}'$.
According to Eq.~(\ref{prob_unconstrained}) and the Metropolis rule, this move is accepted with a probability
$P_\mathrm{acc}(\mathbf{s}\to \mathbf{s}')=\min\left( 1, e^{-\beta \Delta\mathcal{U}}\right)$.
Similarly, the insertion of a particle at a random position and the removal of a random particle are accepted with respective probabilities
\begin{eqnarray}
P_\mathrm{acc}(N\to N+1)&=&\min\bigg[1, \frac{ V e^{-\beta(\Delta\mathcal{U}-\mu)}}{\Lambda^3(N+1)} \bigg],
\label{acceptance_particle_insertion} \\
P_\mathrm{acc}(N\to N-1)&=&\min\big[1, V^{-1}\Lambda^{3}Ne^{  -\beta( \Delta\mathcal{U} + \mu)} \big].
\label{acceptance_particle_removal}
\end{eqnarray}
Finally, trial moves that attempt to change the volume from $V$ to $V'$ are accepted with probability
\begin{equation}
P_\mathrm{acc}(V\to V')=\min \bigg\{1,\frac { e^{ -\beta [\Delta\mathcal{U} + P(V'-V)] } }{ e^{-N\ln(V'/V)}  } \bigg\}. 
\label{acceptance_volume_desplacement}
\end{equation}
Equations (\ref{acceptance_particle_insertion}) and (\ref{acceptance_particle_removal}) correspond to particle insertion and removal acceptance probabilities in the grand canonical ensemble, while Eq.~(\ref{acceptance_volume_desplacement}) is the acceptance probability for volume changes in the isothermal-isobaric ensemble~\cite{Frenkel}.
Therefore, a consistent MC algorithm for simulations in the unconstrained ensemble can be obtained as a simple combination of the usual algorithms for these two ensembles. 
In the Supplemental Material~\cite{Supplemental} we describe the details of the procedure adopted to decide which kind of move is implemented at each MC step.
We now illustrate two applications of the method.

As a first example, consider a system with spatially constant, repulsive interactions $\phi(\mathbf{r}_i,\mathbf{r}_j)=\varepsilon$, for which the potential energy is $\mathcal{U}= \sum_{i>j}^N\phi(\mathbf{r}_i,\mathbf{r}_j)=\frac{1}{2}\varepsilon N(N-1)$, where $\varepsilon>0$ is a coupling constant. 
This system is nonadditive because interactions are long-ranged, regardless of its size.
Although interactions not depending on the interparticle distances may be difficult to justify physically, with this simple model we will show that repulsive interactions in nonadditive systems can withstand completely open conditions resulting in states of thermodynamic equilibrium.
For the present case, the canonical partition function is $Q(N,V,T)=V^N e^{-\frac{1}{2}\beta \varepsilon N(N-1)} / (\Lambda^{3N}N!)$, from which the unconstrained partition function~(\ref{unconstrained_partition_fucntion}) becomes
\begin{equation}
\Upsilon(\mu,P,T)=\sum_{N=0}^\infty e^{[\beta\mu -\ln(\beta P \Lambda^3)+\frac{1}{2}\beta\varepsilon ] N-\frac{1}{2}\beta \varepsilon N^2} .
\label{Upsilon_long-range}
\end{equation}
Since $\varepsilon>0$, the probability of observing the system with $N$ particles, given by the exponential in
Eq.~(\ref{Upsilon_long-range}), has a maximum at 
\begin{equation}
\bar{N}(\mu,P,T)=\varepsilon^{-1}k_BT[\beta\mu - \ln(\beta P \Lambda^3)] +1/2
\label{Nbar}
\end{equation}
with the control parameters taken such that $\bar{N}>0$.
Introducing $T^*= k_BT/\varepsilon$ and $x=N/T^*$, we rewrite the partition function~(\ref{Upsilon_long-range}) as $\Upsilon= e^{\frac{1}{2} T^*\bar{x}^2} \sum_{N} e^{- \frac{1}{2}T^*( x - \bar{x} )^2}$,
where $\bar{x}=\bar{N}/T^*$.
Hence, the distribution has a sharp peak around $\bar{x}$ in the limit $T^*\to\infty$, requiring $\bar{N}\to\infty$ with $\bar{x}$ fixed. Since $\mathscr{E}=-k_BT\ln\Upsilon$, we have $\mathscr{E}=-\frac{1}{2}\varepsilon\bar{N}^2$ in this limit, which displays the dependence of the replica energy on $\mu$, $P$,
and $T$ through Eq.~(\ref{Nbar}).
From this relation, it follows that the average volume $\bar{V}=\partial \mathscr{E}/\partial P$ satisfies $P\bar{V}=\bar{N}k_BT$, just like for an ideal gas.
The basic feature introduced by the interactions, however, is that the size of the system can be controlled with independent $\mu$, $P$, and $T$, which is impossible for a macroscopic ideal gas.
Moreover, we have shown that a large $\bar{N}$ can be realized for temperatures $k_BT\gg \varepsilon$, which can be seen as a weak coupling limit for small $\varepsilon$.
In order to get a finite density $\bar{N}\ell^3/\bar{V}=P^*/T^*$ in the thermodynamic limit, the reduced pressure $P^*=P\ell^3/\varepsilon$ must be large as well, where $\ell$ is an arbitrary unit of length.
Accordingly, the reduced chemical potential $\mu^*=\mu/\varepsilon - T^*\ln(\Lambda^3/\ell^3)$ has to be taken of the same order as $T^*$ and $P^*$, since, from Eq.~(\ref{Nbar}), we see that $\bar{N}=\mu^* - T^*\ln(P^*/T^*) +1/2$.

\begin{figure}
\includegraphics[width=\columnwidth]{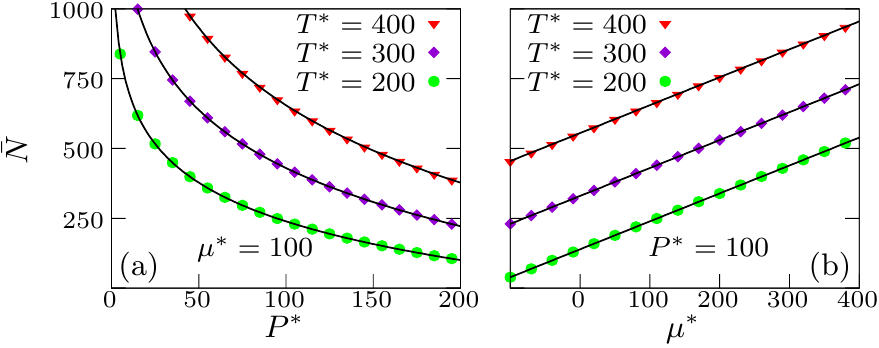}
\caption{MC simulations of a system with spatially constant, long-range repulsion in the unconstrained ensemble: fixing the chemical
potential and varying the applied pressure in (a) and fixing the applied pressure and varying the chemical potential in (b). Solid curves in (a) and (b) are obtained from the expression $\bar{N}=\mu^* - T^*\ln(P^*/T^*) +1/2$.}
\label{fig_repulsion}
\end{figure}

The average number of particles is known analytically in this simple model. To test the proposed MC scheme, we perform simulations in the unconstrained ensemble for this system taking $\mu^*$, $P^*$, and $T^*$ as control parameters.
In this case, the potential energy remains constant for random particle displacements and variations of volume, while  $\Delta\mathcal{U} =\varepsilon N$ and $\Delta\mathcal{U} =\varepsilon(1-N)$ for the insertion and removal of a particle, respectively.
Particle displacements are only rejected when the particle leaves the simulation box (we do not use periodic boundary conditions here) and the remaining acceptance probabilities can be easily obtained from Eqs.~(\ref{acceptance_particle_insertion})-(\ref{acceptance_volume_desplacement}), whose explicit
expressions are given in the Supplemental Material~\cite{Supplemental}.
The simulations are shown in Fig.~\ref{fig_repulsion} for fixed $\mu^*$ while $P^*$ is varied, and varying $\mu^*$ with fixed $P^*$, in both cases for different values of $T^*$.
Solid curves in this figure correspond to the analytical expression for $\bar{N}$. 

Despite the fact that completely open conditions can be achieved by assuming the system coupled to two reservoirs, controlling $\mu$, $P$, and $T$ independently may be challenging in practice.
In what follows we describe an example in which completely open conditions can be readily conceived and which shares key similarities with the previous example, even though this is not evident at first sight.
We consider a system of $N$ hard spheres of diameter $\sigma$ at temperature $T$ confined between parallel plates of area $A$ and gap thickness $H$, so the volume of the system is $V=AH$. 
The dimensions of the plates are constant and much larger than the particle diameter $\sigma$, while $H$ is comparable to $\sigma$. 
Neglecting hard-core interactions between particles but keeping $A(H-\sigma)$ as the available volume due to particle-plate interactions, the system behaves as an ideal gas with free energy $F$ given by~\cite{Schmidt_1997} $\beta F/N=\ln\{N\Lambda^3/[A(H-\sigma)]\}-1$.
In this situation one can define a lateral pressure $P_\mathrm{lat}=-H^{-1}\partial F/\partial A$ (acting in directions parallel to the plates) and a transversal pressure $P=-A^{-1}\partial F/\partial H$ (acting in the direction perpendicular to the plates), which can be shown to be
related to each other through~\cite{Schmidt_1997}
\begin{equation}
P= H P_\mathrm{lat}/(H-\sigma).
\label{lat_pressure}
\end{equation}
Clearly, $P\to P_\mathrm{lat}$ as $H\to\infty$, so they are no longer independent in this limit.
In addition, the chemical potential is given by $\mu=\partial F/\partial N$ and can be written as
\begin{equation}
\mu=k_BT\ln\{N\Lambda^3/[A(H-\sigma)]\}.
\label{mu_spheres}
\end{equation}
Thus, the canonical replica energy $\mathscr{E}=F+PV-\mu N$ takes the form $\mathscr{E}= H^2 P_\mathrm{lat}A/(H-\sigma)-Nk_BT$.
When $H$ and $\sigma$ are comparable, the system is nonadditive because $\mathscr{E}\neq 0$.
In the limit $H\to\infty$, $P_\mathrm{lat}AH=PV\to Nk_BT$, so $\mathscr{E}\to0$ and additivity is recovered.

Assume now that, in this idealized approximation, the canonical and unconstrained ensembles are equivalent and let $H$ and $N$ be fluctuating quantities with averages $\bar{H}$ and $\bar{N}$, respectively. Ensemble equivalence does not hold, in general, when interactions are included.
We suppose that the plates are surrounded by a fluid, acting as a reservoir in equilibrium with the system, with temperature $T$ and chemical potential $\mu_r$ which therefore fixes its pressure $P_r$ (as in a narrow pore with
slit geometry~\cite{Schoen_1994,Evans_1990}).
Since the system is laterally open, equilibrium requires that $\mu=\mu_r$ and $P_\mathrm{lat}=P_r$, while the transversal pressure $P$ can be externally controlled regardless of the value of $P_r$, provided the gap thickness is not too large. 
One possible way to control $P$ is by applying weights to the plates, so these exert on the system a transversal pressure larger than the pressure $P_r$ of the surrounding fluid.
Thus, $\mu$, $P$, and $T$ are independent and define the state of the system.
A reason for this is that the pressure $P$ is not an intensive quantity, since it depends on the size of the system as can be seen from Eq.~(\ref{lat_pressure}).
Although the above arguments presume an ideal system, they capture in first approximation the behavior of hard-sphere systems including the interactions, as we test below with simulations in the unconstrained ensemble.

We now concentrate on a more realistic treatment of the confined hard-sphere system by means of the MC method.
In the simulations, lengths are measured in units of $\sigma$ and energies in units of $k_BT$.
Accordingly, we introduce the reduced pressure $P^*= \beta P\sigma^3$ and chemical potential $\mu^*=\beta \mu -\ln (\Lambda^3/\sigma^3)$, while the temperature is just a scaling factor which will be assumed constant.
Periodic boundary conditions are implemented at the limits of the box in the transversal directions, while hard walls are assumed in the direction perpendicular to the plates.
All MC moves in the unconstrained ensemble, consisting of particle displacements, insertion and removal of particles, and variations of the box length $H$ at constant $A$, are rejected if they lead to an overlap between particles, between particles and the plates, and between the plates.
When there is no overlap, particle displacements are always accepted and the remaining acceptance probabilities can be obtained by direct substitution from Eqs.~(\ref{acceptance_particle_insertion})-(\ref{acceptance_volume_desplacement}) taking $V=AH$ and a vanishing potential energy (see the Supplemental Material~\cite{Supplemental}).
Scaling of particles coordinates for volume changes are performed in the direction perpendicular to the plates only.
Here we further consider that the fluid surrounding the plates consists of hard spheres described by the Carnahan-Starling equation of state~\cite{Carnahan_1969}, for which the pressure is 
$P_r^*(\eta_r)=(6\eta_r/\pi )(1+\eta_r+\eta_r^2-\eta_r^3)/(1-\eta_r)^3$
in our reduced units, where $\eta_r $ is the packing fraction in this fluid.
The corresponding chemical potential~\cite{Lee_1995} reads
$\mu_r^*(\eta_r)=\ln (6\eta_r/\pi )+ (8\eta_r -9\eta_r^2+3\eta_r^3)/(1-\eta_r)^3$.
We emphasize that this characterization of the reservoir serves only to evaluate the simulations, since the actual packing fraction of the system is determined by the control parameters $\mu^*$ and $P^*$. 

\begin{figure}
\includegraphics[width=\columnwidth]{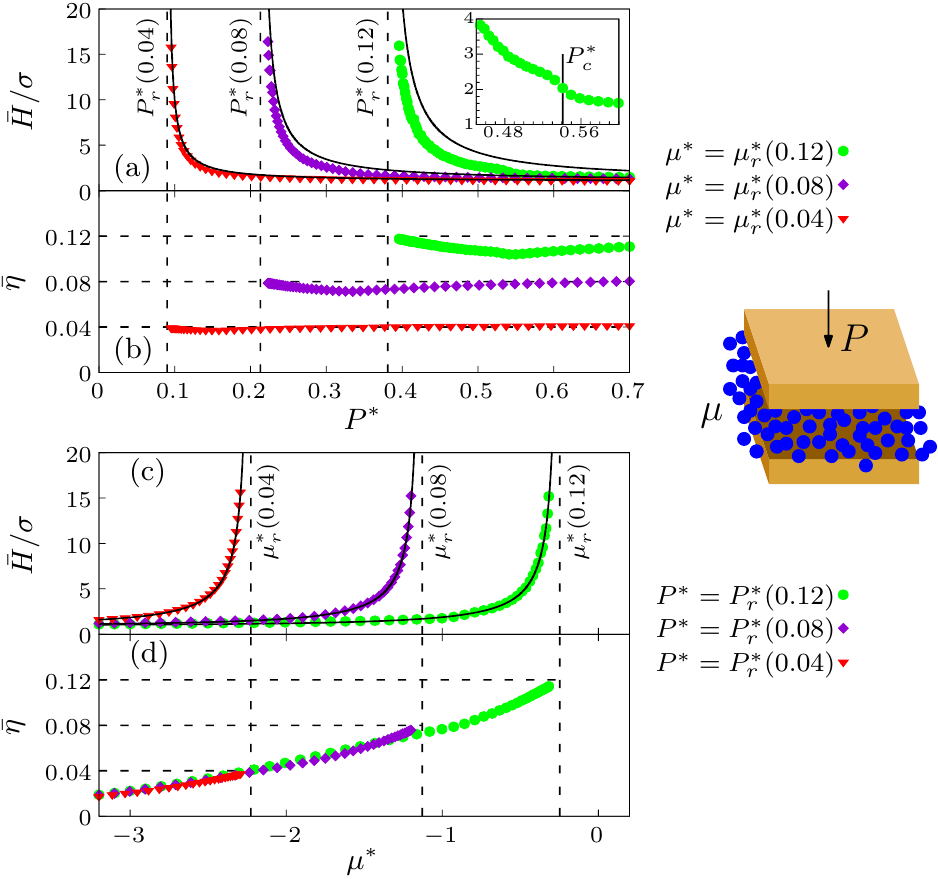}
\caption{MC simulations of confined hard spheres in the unconstrained ensemble: fixing the chemical potential and varying the applied pressure in (a) and (b), and fixing the applied pressure and varying the chemical potential in (c) and (d). Solid lines in (a) correspond to the approximation $P^*=\bar{H}P^*_r(\eta_r)/(\bar{H}-\sigma)$, while in (c) to the approximation $\mu^*=\mu_r^*(\eta_0) -\ln\left[ \bar{H} /(\bar{H}-\sigma) \right]$.}
\label{fig_confined}
\end{figure}

To be able to compare the simulations with the analytical approximation discussed above, we restrict ourselves to small $\eta_r$.
We first choose $\eta_r=0.04$, $0.08$, and $0.12$ and fix the chemical potential of the system to $\mu^*=\mu_r^*(\eta_r)$, meanwhile an external pressure $P^*$ is applied on the plates (their area is fixed to $A/\sigma^2=20^2$).
In Figs.~\ref{fig_confined}(a) and \ref{fig_confined}(b), we show the average gap thickness $\bar{H}$ and the packing fraction $\bar{\eta} = \frac{\pi}{6}\sigma^3 \langle N/V\rangle $ as a function of $P^*$.
We observe that $\bar{H}$ becomes large and approaches a \textit{macroscopic limit} when the applied pressure
approaches the pressure $P_r^*(\eta_r)$ of the reservoir.
In this limit, the freedom to control $\mu^*$ and $P^*$ independently at fixed temperature is lost, as expected.
Solid lines in Fig.~\ref{fig_confined}(a) correspond to the approximation $P^*=\bar{H}P^*_r(\eta_r)/(\bar{H}-\sigma)$ given by Eq.~(\ref{lat_pressure}). 
Furthermore, a smooth kink around $\bar{H}\approx 2\sigma$ is observed at $P^*=P^*_c\approx 0.54$ for the case $\eta_r=0.12$, which is enlarged in the inset of Fig.~\ref{fig_confined}(a). 
For decreasing pressures $P^*>P^*_c$, the gap slowly increases and the system is quasi-bidimensional because configurations with two particles aligned in the direction perpendicular to the plates are not realized.
Fluctuations of the gap size at $P^*_c$ allow such configurations to be realized and the gap grows faster for decreasing pressures $P^*<P^*_c$.
The average packing fraction consequently decreases for decreasing pressures $P^*>P^*_c$.
This behavior is also observed for $\eta_r=0.04$ and $0.08$, but it is less pronounced in the plots.
We next consider a situation in which $P^*$ is fixed to a value $P^*=P_r^*(\eta_0)$ for some packing fraction $\eta_0$.
Then, the chemical potential of the reservoir is varied (by changing $\eta_r$), so $\mu^*=\mu_r^*(\eta_r)$ also changes.
In other words, we control $\mu^*$ keeping $P^*$ constant.
The results of the simulations are shown in Figs.~\ref{fig_confined}(c) and \ref{fig_confined}(d) for $\eta_0=0.04$, $0.08$, and $0.12$.
When $\mu^*$ approaches $\mu_r^*(\eta_0)$, the average thickness $\bar{H}$ approaches the macroscopic limit.
Solid lines in Fig.~\ref{fig_confined}(c) correspond to the approximation
$\mu^*=\mu_r^*(\eta_0) -\ln\left[ \bar{H} /(\bar{H}-\sigma) \right]$, obtained by combining Eqs.~(\ref{lat_pressure}) and (\ref{mu_spheres}).
Therefore, we have shown that chemical potential and pressure can be independently controlled in this system at fixed temperature.
We emphasize that the hard-core repulsion makes the system nonadditive for gap sizes comparable to $\sigma$, since $\mathscr{E}\neq0$ in this case.
Similar to the previous example, a key point here is that repulsive interactions maintain equilibrium states under completely open conditions in a regime in which the system is nonadditive.

When restricting to the particular situation of a confined geometry as in the example above, we highlight that the MC scheme presented here is similar to the approach introduced in Ref.~\cite{Schoen_1994} describing the grand isostress ensemble. In that case, the replica energy $\mathscr{E}$ plays the role of the grand
isostress potential considered there.

In conclusion, we have shown that a consistent MC scheme for simulations in the unconstrained ensemble can be obtained by combining the algorithms of the grand canonical and isothermal-isobaric ensembles.
This scheme applies to nonadditive systems in which chemical potential, pressure, and temperature can be controlled independently.
We have also shown with some examples that repulsive interactions can hold a nonadditive system in equilibrium under completely open conditions.
While the implementation of the proposed scheme does not present any further difficulties other than those inherently associated with the isobaric-isothermal and grand canonical ensembles, we remark that nonadditivity is required to observe equilibrium states in the unconstrained ensemble.
The proposed method paves the way for new developments in simulations of systems that exchange heat, work, and matter with their environment.
Beyond long-range interacting systems and confined systems as considered here, such environmental conditions can be relevant, e.g., in small self-assembled aggregates~\cite{Sciortino_2004,Mossa_2004,Santos_2017} whose structures are stabilized by the competition of repulsive and attractive interactions.

\begin{acknowledgments}
S.\,R. thanks Peter Sollich for discussions and for suggesting Refs.~\cite{Orkoulas_2009,Wilding_2013}.
A.\,C. acknowledges financial support from INFN (Istituto Nazionale di Fisica Nucleare) through the projects DYNSYSMATH and ENESMA.
L.\,C., P.\,D.\,C., and S.\,R. acknowledge partial support from the MIUR-PRIN2017 Project No. 201798CZL, ``Coarse-grained description for non-equilibrium systems and transport phenomena (CO-NEST)''.
J.\,M.\,R. acknowledges financial support from the MICIU of the Spanish Government under Grant No. PGC2018-098373-B-I00.
This project has received funding from the European Union’s Horizon 2020 research and innovation
programme under the Marie Sklodowska-Curie grant agreement No 892718.
\end{acknowledgments}

\vspace{10mm}
\hrule

\setcounter{equation}{0}
\begin{center}
\textbf{\large Supplemental Material}

\end{center}

\section{Replica energy and thermodynamic relations}

Here we explicitly integrate the energy balance equation for
the ensemble, following the arguments in Ref.~\cite{Latella_2015} (see also~\cite{Hill}). As stated in Eq.~(1) of the main text, for a collection of
$\mathscr{N}$ independent replicas with total energy, entropy, volume, and number of particles $E_t=\mathscr{N} E$, $S_t=\mathscr{N} S$,
$V_t=\mathscr{N} V$, and $N_t=\mathscr{N} N$, respectively, one has the general expression
\begin{equation*}
dE_t=TdS_t-PdV_t+\mu dN_t+\mathscr{E}d\mathscr{N}.
\end{equation*}
Keeping all single-system properties constant, one obtains $dE_t=Ed\mathscr{N}$, $TdS_t=TSd\mathscr{N}$, $PdV_t=PVd\mathscr{N}$, and
$\mu dN_t= \mu Nd\mathscr{N}$, so that
\begin{equation*}
Ed\mathscr{N}=TS d\mathscr{N}-PV d\mathscr{N}+\mu N d\mathscr{N}+\mathscr{E}d\mathscr{N}.
\end{equation*}
Now the integration is to be performed on the number of replicas only. Integrating between $0$ and $\mathscr{N}$ yields
\begin{equation*}
E\mathscr{N}=TS \mathscr{N}-PV \mathscr{N}+\mu N \mathscr{N}+\mathscr{E}\mathscr{N},
\end{equation*}
and therefore we arrive at $\mathscr{E}= E-TS+PV-\mu N$, which is always valid, in particular for any $N$.

\section{Probability density in the unconstrained ensemble}

Here we present a derivation of the probability density in the unconstrained ensemble describing the system in a particular $N$-particle configuration, given in Eq.~(3) of the main text. We follow Ref.~\cite{Frenkel}, extending the standard arguments for other ensembles.

The system has $N$ particles, volume $V$, and temperature $T$, and we recall that its canonical partition function takes the form
\begin{equation}
Q(N,V,T)=\frac{V^N}{\Lambda^{3N}N!}\int_0^1\cdots\int_0^1 d\mathbf{s}^N 
e^{-\beta\mathcal{U}(\mathbf{s}^N;V)},
\label{Q_canonical_supp}
\end{equation}
where $\beta =1/k_BT$, $\Lambda$ is the de Broglie thermal wavelength, and $\mathcal{U}(\mathbf{s}^N;V)$ is the potential energy, $k_B$ being the Boltzmann constant.
In addition, reservoirs $a$ and $b$ are assumed to be ideal gases and their canonical partition functions are given by
\begin{equation}
Q_a(N_a-N,V_a,T)=\frac{V_a^{N_a-N}}{\Lambda^{3(N_a-N)}(N_a-N)!},
\end{equation}
and
\begin{equation}
Q_b(N_b,V_b-V,T)=\frac{(V_b-V)^{N_b}}{\Lambda^{3N_b}N_b!},
\end{equation}
respectively, where reservoir $a$ has $N_a-N$ particles and volume $V_a$ and reservoir $b$ has $N_b$ particles and volume $V_b-V$. 

We now consider the limit in which the reservoirs are infinite. Let us first focus on reservoir $a$. We take $N_a\to\infty$, $V_a\to\infty$ with $N_a/V_a=\rho_a$, and use the limit $N_a/N\to\infty$ in the partition function such that
\begin{equation}
Q_a=e^{-N_a\ln(\Lambda^{3}\rho_a) +N_a +N\ln(\Lambda^3\rho_a)}.
\end{equation}
In this limit, a change in the number of particles of the system does not change the chemical potential $\mu$ of reservoir $a$.
Since the reservoir is an ideal gas, its chemical potential is given by $\mu=k_BT\ln(\Lambda^3\rho_a)$ and hence
\begin{equation}
Q_a=e^{-N_a\ln(\Lambda^{3}\rho_a) +N_a} e^{ \beta \mu N}.
\label{Q_a}
\end{equation} 
For reservoir $b$, we take $N_b\to\infty$, $V_b\to\infty$ with $N_b/V_b=\rho_b$.
In the limit $V/V_b\to 0$, we can write
\begin{equation}
(V_b-V)^{N_b} =V_b^{N_b}[1-(V/V_b)]^{N_b}\to V_b^{N_b}e^{-N_b V/V_b }. 
\end{equation}
In this limit, a change in the volume of the system does not change the pressure $P$ of reservoir $b$. Since this reservoir is an ideal gas as well, the density $\rho_b=N_b/V_b$ can be written as $\rho_b= \beta P$ and hence, $(V_b-V)^{N_b}\to V_b^{N_b}e^{-\beta P V }$. Using these formulas, the partition function $Q_b$ becomes
\begin{equation}
Q_b=\frac{V_b^{N_b}}{\Lambda^{3N_b}N_b!}e^{-\beta P V }.
\label{Q_b}
\end{equation}
Taking into account (\ref{Q_a}) and (\ref{Q_b}) to compute the partition function of the total system $Q^\mathrm{tot}=Q_aQQ_b$, in the limit $N_a,V_b\to\infty$ the probability density $\mathcal{P}(N,V)=Q^\mathrm{tot}/\sum_{N=0}^{N_a}\int_0^{V_b}dVQ^\mathrm{tot}$ becomes
\begin{equation}
\mathcal{P}(N,V)=\frac{ \beta P e^{\beta\mu N-\beta PV}Q(N,V,T)}{\Upsilon(\mu,P,T)} ,
\label{prob_unconstrained_supp}
\end{equation}
where
\begin{equation}
\Upsilon(\mu,P,T)=\beta P \sum_{N=0}^\infty e^{\beta\mu N}\int_0^\infty dV e^{-\beta PV}Q(N,V,T)
\label{unconstrained_partition_fucntion_supp}
\end{equation}
is the unconstrained partition function of the system. Here the factor $\beta P$ is included to make $\Upsilon$ a dimensionless quantity, as it is usually done for the isothermal-isobaric partition function~\cite{Frenkel,Hill_SM,Hill}.
Using the canonical partition function (\ref{Q_canonical_supp}), the probability density (\ref{prob_unconstrained_supp}) can be rewritten as
\begin{equation}
\mathcal{P}(N,V)=\frac{ \beta P e^{\beta\mu N-\beta PV}V^N}{\Upsilon(\mu,P,T)\Lambda^{3N}N!} 
\int d\mathbf{s}^N 
e^{-\beta\mathcal{U}(\mathbf{s}^N;V)}.
\label{prob_unconstrained_2}
\end{equation}
From this expression one directly gets the probability density $\mathcal{P}(N,V;\mathbf{s}^N)$ in a particular $N$-particle configuration, as given in Eq.~(3) of the main text.

\section{Simulation details}

In this section we give more details about the simulations of the examples considered in the main text and the explicit expressions of the acceptance probabilities.

The number of Monte Carlo (MC) moves in a cycle is defined by $m=N_\mathrm{av}+N_\mathrm{ex}+1$, where $N_\mathrm{av}$ and $N_\mathrm{ex}$ are fixed integers. To implement the algorithm, we generate a random integer $R$ such that $1\leq R\leq m$ and attempt a particle displacement if $R\leq N_\mathrm{av}$, a volume change if $R=N_\mathrm{av}+1$, and a particle exchange with the reservoir (insertion or removal with the same probability) otherwise. In this way, on average, per cycle the algorithm performs $N_\mathrm{av}$ particle displacements, $N_\mathrm{ex}$ particle exchanges, and one volume change. During all the simulations we set $N_\mathrm{ex}=1$.

Before the production run, a calibration stage is performed in the simulations followed by a thermalization stage~\cite{Miller_2000}. In the calibration stage, the maximum particle displacement and maximum volume variation are periodically updated to achieve an acceptance ratio of about $50\%$. Also, in this stage we periodically set $N_\mathrm{av}=N$ to enforce that $N_\mathrm{av}$ is close to the average number of particles which is a priori unknown. In the subsequent thermalization stage, simulations are carried out with all parameters fixed, and the average number of particles $\bar{N}$ is computed. At the end of this stage, we set $N_\mathrm{av}=\bar{N}$. Finally, keeping all parameters fixed, we compute the averages in the production run, including the final value of the average number of particles $\bar{N}$. For each point in the plots of the simulations shown in the main text, the number of MC moves has been set to $10^6$ during calibration, $10^9$ during thermalization, and $3\times10^9$ or $8\times10^9$ in the production stage.

The acceptance probabilities for insertion and removal of particles and for volume changes in the unconstrained ensemble are obtained from expressions (5)-(7) of the main text. In the first example (system with spatially constant, repulsive interactions), using the reduced control parameters $\mu^*$, $P^*$, and $T^*$ introduced in the main text, these equations become
\begin{align}
P_\mathrm{acc}(N\to N+1)&=
\min\bigg[1, \frac{ V^* e^{(\mu^*-N)/T^* }}{(N+1)} \bigg],\\
P_\mathrm{acc}(N\to N-1)&=
\min\left[1, \frac{N e^{-(\mu^*+1-N)/T^*} }{V^*}\right],\\
P_\mathrm{acc}(V\to V')&=
\min \left[1, 
\frac { e^{ -P^*({V^*}'-V^*)/T^* } }
{ e^{-N\ln({V^*}'/V^*)}  } \right],
\end{align}
where $V^*=V/\ell^3$ and ${V^*}'=V'/\ell^3$, $\ell$ being an arbitrary length unit.
For the second example (confined hard spheres), using the reduced control parameters $\mu^*$ and $P^*$ defined for this case in the main text, when there is no overlap the acceptance probabilities take the form
\begin{align}
P_\mathrm{acc}(N\to N+1)&=
\min\bigg[1, \frac{ A^*H^* e^{\mu^*}}{(N+1)} \bigg],\\
P_\mathrm{acc}(N\to N-1)&=
\min\left[1, \frac{Ne^{  -\mu^*}}{A^*H^*}\right],\\
P_\mathrm{acc}(H\to H')&=
\min \bigg\{1, 
\frac { e^{ -P^*A^*({H^*}'-H^*) } }
{ e^{-N\ln({H^*}'/H^*)}  } \bigg\},
\end{align}
where $A^*=A/\sigma^2$, $H^*=H/\sigma$ and ${H^*}'=H'/\sigma$, $\sigma$ being the particle diameter. We stress that in this case volume changes are obtained by performing variations of the gap thickness from $H$ to $H'$ at constant area $A$.



\begin{thebibliography}{99}
\bibitem{footnote}By confined system we mean a system in which one of the typical lengths defining its size, say $L_x$, is such that $L_x/r_0$ is not much larger than $1$, where $r_0$ is the range of the intermolecular interaction.
\end{thebibliography}
\end{document}